\documentclass[prd,preprint,showpacs,amsmath,amssymb,floatfix,10pt]{revtex4}
\usepackage{graphicx,color,dcolumn,booktabs}
\usepackage{longtable,lscape}
\usepackage{tabularx}
\usepackage{txfonts}
\usepackage{amssymb}
\usepackage{mathptmx}

\begin{document}
\preprint{IPMU-10-0117}
\title{Comments on $\left|\frac{ V_{ub}}{V_{cb}}\right|$ and $|V_{ub}|$ from Non-leptonic $B$ Decays \\
within the Perturbative QCD Approach}

\author{C.~S. Kim\footnote{Email: cskim@yonsei.ac.kr}}
\affiliation{Department of Physics and IPAP, Yonsei University,
Seoul 120-479, Korea\\
IPMU, University of Tokyo, Kashiwa, Chiba 277-8568, Japan }
\author{Ying Li\footnote{Email: liying@ytu.edu.cn }}
\affiliation{Department of Physics, Yantai University, Yantai
264-005, China \\Department of Physics and IPAP, Yonsei University,
Seoul 120-479, Korea }%

\begin{abstract}

\noindent We revisited the extracting  $\left|V_{ub}/V_{cb}\right|$ and
$|V_{ub}|$ through calculating the ratios   ${\cal B}(B^0\to D_s^{(*)+}(\pi^-,\rho^-)/
{\cal B}(B^0\to D_s^{(*)+} D^{(*)-})$ in the perturbative QCD approach, which is regarded
as an effective theory in dealing with non-leptonic charmed $B$ decays.  Utilizing this approach,
we could calculate the form factors effectively as well as non-factorizable and
annihilation contributions. Within the updated distribution amplitudes and the latest experimental
data, we get  $\left|{V_{ub}}/{V_{cb}}\right|=0.083\pm0.007$, which favors a bit smaller $|V_{ub}|$
compared with the averaged PDG value, but agrees well the exclusively measured values.
Furthermore, we predict the branching ratio of $\bar B^0 \to D_s^-\rho^+$ $\sim$
$(2.7\pm 1.2)\times 10^{-5}$, which could be measured in $B$ factories near future. In our calculation,
the major uncertainty is due to our poor knowledge of heavy meson wave functions.
We also comment that it is not trivial to generalize this approach to $B_s$ system, primarily because
both $\bar B_s^0 $ and $B_s^0$ can decay into  the same final states $D_s^+K^-$ and $D_s^+D_s^-$.

\end{abstract}

\maketitle

The precise measurement of the Cabibbo-Kobayashi-Maskawa (CKM)
matrix elements and the three related  angles, namely $\alpha$,
$\beta$ and $\gamma$, is one of  the most important topics in flavor
physics \cite{Isidori:2008qp}. Through measurements of those observables
we can test the standard model (SM) stringently and seek for a
possible new physics beyond the SM. For instance, nonzero $|V_{ub}|$
plays critical role in explaining the CP violation in $B$ and $K$ decays.
Furthermore, the ratio between $|V_{ub}|$ and $|V_{cb}|$ is
very important as it is proportional the length of the side of
the unitary triangle opposite to the well-measured angle $\beta$. Since
the end of last century, the obvious example of the progress was
made in measuring the ratio $\left| V_{ub}/V_{cb}\right|$ and
$|V_{ub}|$ as acutely as we can \cite{Charles:2004jd}.

In both experimental and theoretical sides, the  determinations of
$|V_{ub}|$ are mostly on semi-leptonic $B$ decays, involving
inclusive $B \to X_u l\bar \nu$ decay and exclusive  $B \to \pi l
\bar \nu$ decay. However, both of them suffer from large
uncertainties due to the unknown non-perturbative dynamics. For
inclusive $B \to X_u l\bar \nu$ decay, in principle, it is the most
appreciated way to extract $|V_{ub}|$. In practice, however, since $|V_{cb}|
\gg |V_{ub}|$, the total width of $B \to X_u l\bar \nu$ cannot be measured due to the large
charm background from $B \to X_c l \bar \nu$ decays. In most regions
of phase space, where the charm background is kinematically
forbidden, the hadronic physics enters via unknown non-perturbative
functions, so-called shape functions. Although  there is only one
shape function at leading order in $\Lambda_{QCD}/m_b$, which can be
extracted from the photon energy spectrum in $B\to X_s \gamma$, the
subleading shape functions are modeled in the current calculations
\cite{Lee:2008vs,Lange:2005yw}. For the exclusive decay, the
determination of $|V_{ub}|$ requires a theoretical calculation of
the form factors using non-perturbative methods such as Lattice QCD
\cite{Dalgic:2006dt,Okamoto:2004xg}, QCD sum rules
\cite{Ball:2004ye} or light front approach \cite{Chen:2009qk}, all of which have systematic uncertainties that
are hard to be quantified. The $B \to \pi l \bar \nu$ branching ratio is
now well known to $6\%$. Unquenched lattice QCD calculations of the $B
\to \pi l \bar \nu$ form factor for $q^2 > 16 \mathrm{GeV}^2$ are
available and yield \cite{Dalgic:2006dt,Okamoto:2004xg}
\begin{eqnarray}|V_{ub}| = (3.4 \pm 0.2^{+0.6}_{-0.4}) \times 10^{-3}.\end{eqnarray}
Light-cone QCD sum rules are applicable for $q^2 < 14 \mathrm{GeV}^2$ and yield similar
results \cite{Ball:2004ye}. The theoretical uncertainties in
extracting $|V_{ub}|$ from inclusive and exclusive decays are very much
different. In PDG  \cite{Amsler:2008zzb},
the values obtained from inclusive and exclusive determinations are
\begin{eqnarray}
|V_{ub}| &=& (4.12 \pm 0.43) \times 10^{-3}
, ~~~~~(\mathrm{inclusive})\label{vubexp1}\\
 &=& (3.5 ^{+ 0.6}_{-0.5} )  \times 10^{-3}, ~~~~~~~~~~~~(\mathrm{exclusive})\label{vubexp2}.
\end{eqnarray}
Above two determinations are independent each other, and the dominant
uncertainties are on multiplicative factors. If we weight the two
values by their relative errors and treat the uncertainties as
Gaussian,  we find \cite{Amsler:2008zzb}
\begin{eqnarray}\label{vubexp3}
|V_{ub}| = (3.93 \pm 0.36) \times 10^{-3},~~~~~~~~~~~~({\rm combined})
\end{eqnarray}
whose determination is dominated by the inclusive measurement. Given the
theoretical and experimental progress, it will be interesting to see
how the inclusive and exclusive determinations develop.

Although the complexity of QCD and weak dynamics  had prevented us
from measuring CKM elements through non-leptonic decays, many methods have been explored in determining
$|V_{ub}|$ though analyzing  decays of $B$ to exclusive non-leptonic
two meson final states. In Refs. \cite{Koide:1989hv,Kim:2000cv}, the authors had
pointed out that we could extract the $|V_{ub}|/|V_{cb}|$ and $|V_{ub}|$ by studying the
ratio of the branching fractions
$${\cal B}(B^0 \to D_s^+ \pi^-)/{\cal B}(B^0 \to D_s^+ D^-).$$
For decays $B^0 \to D_s^+ \pi^-$ and $B^0 \to D_s^+ D^-$, either of decay modes
involves a single isospin component $I=1 (I=1/2)$, so the final
state interaction is not important. Furthermore, the $B^0 \to D_s^+
\pi^-$ is only induced by tree operators; for $B^0 \to D_s^+ D^-$,
although penguin type diagrams will pollute this decay, the tree
operators are still dominant. In Ref. \cite{Kim:2000cv}, the penguin
contribution has been considered in the generalized factorization
scheme, and the decay modes with vector meson  have been also
considered. Within contemporary data, the authors estimated the ratios and
gave the  theoretical upper limit of $\left|{V_{ub}}/{V_{cb}}\right|$. However, in
the generalized factorization approach,  the effective color
number $N_C^{\rm eff}$, a non-physical constant, should be introduced in
order to make up the non-factorizable diagrams' contributions.
In the past ten years, many efforts have been promoted to
estimate the non-factorizabale contributions quantitatively, such as
the QCD fatorization approach (QCDF) \cite{Beneke:1999br}, the perturbative QCD approach (PQCD)
\cite{Keum:2000ph1,Keum:2000ph} and the soft collinear effective theory (SCET) \cite{Bauer:2000yr}.
Due to charmed meson involved,  QCDF and SCET cannot be used here. In
this letter, we reinvestigate the way proposed in
\cite{Kim:2000cv} within the PQCD approach. In Refs.
\cite{Keum:2003js,Li:2008ts,Zou:2009zza,Li:2009xf},  the decay modes with $D$ or
$D_s$ have been also calculated, but their results cannot be used here
directly because different parameters have been used in different
modes. Here we  will adopt the consistent parameter space and the
updated distribution amplitudes. Moreover, through this study we can
cross-check the applicability of the PQCD with charmed meson, which is another
starting point of this work.

For the decay modes here we discuss, the related Hamiltonian and the
corresponding Wilson coefficients  are given in
Ref.~\cite{Buchalla:1995vs}. In the PQCD approach, after the
integration over momenta of positive and negative direction, the
decay amplitude for $B \to M_2 M_3$ can be conceptually written
as
\begin{eqnarray} {\cal A} \sim \int\!\! d x_1 d x_2 d
x_3 b_1 d b_1 b_2 d b_2 b_3 d b_3  \mathrm{Tr} \left [ C(t)
\Phi_{B}(x_1,b_1) \Phi_{M_{2}}(x_2,b_2)\Phi_{M_{3}}(x_3, b_3) H(x_i, b_i, t)
S_t(x_i)\, e^{-S(t)} \right ], \label{eq:a2}
\end{eqnarray}
where $b_i$ is the conjugate space coordinate of intrinsical
transverse momentum of light quark, and $x_i$ is momenta fraction
of light quark  in each meson. $\mathrm{Tr}$ denotes the trace over
Dirac and color indices, $C(t)$ is the Wilson coefficient evaluated
at scale $t$, and  $H(x_i, b_i, t)$ is the hard QCD part, which
can be calculated perturbatively. The function $\Phi_M$ is
the wave function of a meson $M$, and the function $S_t(x_i)$ describes the threshold
re-summation which smears the end-point singularities on $x_i$.
The last term, $e^{-S(t)}$, is the Sudakov form factor which
suppresses the soft dynamics effectively.
 More detailed information on the PQCD can be refereed to
Ref. \cite{Keum:2000ph}. In the PQCD picture, the decay amplitudes of
related decay modes are written as
\begin{eqnarray}\label{eq:amplitude1}
{\cal A}(\bar B^0 \to D_s^- D^+)&=&\frac{G_F}{\sqrt{2}}\Big\{
V_{cb}V^*_{cs}[F_e^{LL}(C_2+\frac{1}{3}C_1)+F_{en}^{LL}(C_1)]-V_{tb}V^*_{ts}
[F_e^{LL}(C_4+\frac{1}{3}C_3+C_{10}+\frac{1}{3}C_9)
+F_{en}^{LL}(C_3+C_9)\nonumber\\
&&\left.+F_e^{SP}(C_6+\frac{1}{3}C_5+C_8+\frac{1}{3}C_7)+F_{en}^{LR}(C_5+C_7)+
F_a^{LL}(C_4+\frac{1}{3}C_3-\frac{1}{2}C_{10}-\frac{1}{6}C_{9})\right.
\nonumber\\
&&+F_{an}^{LL}(C_3-\frac{1}{2}C_9)+F_a^{SP}(C_6+\frac{1}{3}C_5-\frac{1}{2}C_{8}-
\frac{1}{6}C_{7})+F_{an}^{LR}(C_5-\frac{1}{2}C_7)]\Big
\},\\\label{eq:amplitude2}
{\cal A}(\bar B^0 \to D_s^- \pi^+) &=& \frac{G_F} {\sqrt{2}}
V_{ub}V_{cs}^*
\Big\{F_e^{LL}(C_2+\frac{1}{3}C_1)+F_{en}^{LL}(C_1)\Big\}.
\end{eqnarray}
In above formulae, $F_{e(n)}^{LL(R)}$ is the amplitude from (non)-factorizable
emission-diagrams with $(V-A)(V-A)$ (or $(V-A)(V+A)$)
operators. $SP$ stands for the operators $-2(S-P)(S+P)$, which comes
from the Fierz translation of $(V-A)(V+A)$ of different color
structure. Also, $a(n)$ means the contribution from (non)-factorizable
annihilation diagrams. The detailed expression of each $F$ can be found in
Refs. \cite{Li:2008ts,Zou:2009zza,Li:2009xf}, and we would not list them explicitly
here. Note that the  $F_i$'s are different for different decay
modes, as they are mode-dependent. In the calculation, the parameters
we input are the non-perturbative parts, namely the distribution
amplitudes and decay constants. For the light mesons $\pi$ and
$\rho$, the distribution amplitudes are well studied within the
light-cone QCD sum rules, and we will use the latest versions \cite{Ball:2006wn}. However,
for heavy mesons $B$ and $D$, their wave functions and distribution
amplitudes are still less known, even though much efforts have been
made \cite{Grozin:1996pq}. For the $B$ meson distribution amplitude, we
adopt the model \cite{Keum:2000ph1}:
\begin{eqnarray}
\phi_{B}(x,b)=N_{B}x^{2}(1-x)^{2}\exp \left[ -\frac{1}{2} \left(
\frac{xM_{B}}{\omega _{B}}\right) ^{2} -\frac{\omega
_{B}^{2}b^{2}}{2}\right] \label{bw} \;,
\end{eqnarray}
with the shape parameter $\omega_{B}=0.40\pm0.05$ GeV, which has been
tested in many channels such as $B\to \pi\pi, K\pi$
\cite{Keum:2000ph}. The normalization constant $N_{B}$ is related to
the decay constant $f_{B}=190$ MeV~ \cite{Keum:2000ph1}.  As for $D$ or $D_s$ meson, the
distribution amplitude, determined in Ref. \cite{Li:2008ts} by
fitting, is
\begin{eqnarray}\label{Ddis}
\phi_D=\frac{1}{2\sqrt{6}}f_D6x(1-x)\left[1+C_D(1-2x)\right]\exp
\left[ -\frac{\omega^2b^2}{2}\right],
\end{eqnarray}
where $C_D=0.5, \omega=0.1$ for $D$ meson and $C_D=0.4, \omega=0.2$
for $D_s$.
\begin{table}
\begin{center}
\caption{The numerical results of $T$ and $P$ for each decay mode. Theoretical Results are the
predictions of the branching ratios in the PQCD approach. For comparison, we also list the experimental data. }
\begin{tabular}{|c|c|c|c|c|}
  \hline
Mode                                 & $T$ & $P$ & Theoretical Results &
Experimental Data \cite{Amsler:2008zzb}\\ \hline
$ \bar{B}^0\to D_s^-\pi^+    $  &$ 1.56-0.01~i $&$ 0 $&$ (2.6^{+0.9+1.2}_{-0.6-0.8})\times 10^{-5}  $&$ (2.4\pm0.5)\times 10^{-5} $\\
$ \bar{B}^0\to D_s^-\rho^+   $  &$ 1.64-0.04~i $&$ 0 $&$ (3.1^{+1.0+1.3}_{-0.8-1.0})\times 10^{-5}  $&$ <2.4\times 10^{-4}$ \\
$ \bar{B}^0\to D_s^{*-}\pi^+ $  &$ 1.47-0.02~i $&$ 0 $&$ (2.7^{+0.9+1.2}_{-0.6-0.8})\times 10^{-5} $&$ (2.6^{+0.5}_{-0.4})\times 10^{-5} $\\
$ \bar{B}^0\to D_s^-D^+      $  &$ 2.67-0.17~i $&$ 0.23-0.01~i $&$(7.7^{+1.9+4.3}_{-1.4-2.7})\times 10^{-3}  $&$ (7.4\pm 0.7)\times 10^{-3} $\\
$ \bar{B}^0\to D_s^-D^{*+}   $  &$ 2.68-0.17~i $&$ 0.23-0.02~i $&$(7.6^{+1.9+4.3}_{-1.4-2.7})\times 10^{-3} $&$ (8.2\pm 1.1)\times 10^{-3} $\\
$ \bar{B}^0\to D_s^{*-}D^+   $  &$ 2.72-0.21~i $&$ 0.25-0.02~i $&$(7.7^{+1.9+4.3}_{-1.4-2.7})\times 10^{-3}  $&$ (7.6\pm 1.6)\times 10^{-3} $\\
\hline
\end{tabular}

\end{center}
\end{table}

Because of the unitarity of the CKM matrix, we have a relation:
\begin{eqnarray}\label{eq:ckmrelation}
V_{ub}V^*_{us}+V_{cb}V^*_{cs}+V_{tb}V^*_{ts}=0.
\end{eqnarray}
Within the parametrization proposed by Wolfenstein,
$|V_{ub}V^*_{us}|$ is about $\lambda^4$, while $|V_{tb}V^*_{ts}|$
and $|V_{cb}V^*_{cs}|$ are about $\lambda^2$,~ where $\lambda \simeq 0.22$. Therefore,
$|V_{ub}V^*_{us}|$ can be ignored safely, which leads to
$V_{tb}V^*_{ts}=-V_{cb}V^*_{cs}$. For simplicity, Eqs. (\ref{eq:amplitude1}--\ref{eq:amplitude2}) could be re-written as:
\begin{eqnarray}\label{eq:amplitude3}
{\cal A}(\bar B^0 \to  D_s^-D^+)
&=&V_{cb}V^*_{cs}(T_{DD_s}+P_{DD_s}),\\
{\cal A}(\bar B^0 \to D_s^- \pi^+) &=& V_{ub}V_{cs}^* T_{D_s\pi}.
\end{eqnarray}
$T_{DD_s}$, $P_{DD_s}$ and $T_{D_s\pi}$ can be calculated directly
in the PQCD approach, whose central values  are listed in Table. 1.  In
Table. 1, we also list the branching ratios calculated within the PQCD approach
with the parameter values  \cite{Charles:2004jd,Amsler:2008zzb,utfitter}, $i.e.$
$|V_{ub}|=0. 00359$, $|V_{cb}|=0. 041$ and $|V_{cs}|=0.973$. The first error is
from the threshold re-summation, and the second one comes from the
distribution amplitude of heavy meson.
For uncertainties of heavy meson, please look at
Refs. \cite{Li:2008ts,Zou:2009zza,Li:2009xf}. Comparing our theoretical
predictions with experimental data, we find that our predictions are quite close
to the data after combining all kinds of uncertainties, even though the central values are a bit
larger than those of experimental data, which may hint a smaller $|V_{ub}|$.
It should be also noted that our result of $ \bar B^0 \to D_s^- \pi^+$ is smaller than that from
Ref. \cite{Zou:2009zza}, mainly because the different light meson
distribution amplitudes are used.

Following Ref. \cite{Kim:2000cv}, we  define the ratios and estimate the values as,
\begin{eqnarray}
&&R_{\pi/D} \equiv \frac{\mathcal{B}(\bar B^0 \to D_s^-
\pi^+)}{\mathcal{B}(\bar B^0 \to D_s^-D^+ )}=\left|\frac {V_{ub}
}{V_{cb}}\right|^2\left|\frac
{T_{D_s\pi}}{T_{D_sD}+P_{D_sD}}\right|^2
\left(\frac{p_c^\pi}{p_c^D}\right)=\left|\frac {V_{ub}
}{V_{cb}}\right|^2\times(0.44\pm 0.07),\label{eq:ratio1}\\
&&R_{\rho/D} \equiv \frac{\mathcal{B}(\bar B^0 \to D_s^-
\rho^+)}{\mathcal{B}(\bar B^0 \to D_s^-D^+ )}=\left|\frac {V_{ub}
}{V_{cb}}\right|^2\left|\frac
{T_{D_s\rho}}{T_{D_sD}+P_{D_sD}}\right|^2
\left(\frac{p_c^\rho}{p_c^D}\right)=\left|\frac {V_{ub}
}{V_{cb}}\right|^2\times(0.53\pm 0.09),\label{eq:ratio2}\\
&&R_{\pi/D^*} \equiv \frac{\mathcal{B}(\bar B^0 \to D_s^-
\pi^+)}{\mathcal{B}(\bar B^0 \to D_s^-D^{*+} )}=\left|\frac {V_{ub}
}{V_{cb}}\right|^2\left|\frac
{T_{D_s\pi}}{T_{D_sD^*}+P_{D_sD^*}}\right|^2  \left(\frac{p_c^\pi}{p_c^{D*}}\right)=\left|\frac {V_{ub}
}{V_{cb}}\right|^2\times(0.46\pm 0.08),\label{eq:ratio3}\\
&&R_{\rho/D^*} \equiv \frac{\mathcal{B}(\bar B^0 \to D_s^-
\rho^+)}{\mathcal{B}(\bar B^0 \to D_s^-D^{*+} )}=\left|\frac {V_{ub}
}{V_{cb}}\right|^2\left|\frac
{T_{D_s\rho}}{T_{D_sD^*}+P_{D_sD^*}}\right|^2  \left(\frac{p_c^\rho}{p_c^{D*}}\right)=\left|\frac {V_{ub}
}{V_{cb}}\right|^2\times(0.55\pm 0.09),\label{eq:ratio4}\\\label{eq:ratio5}
&&\widetilde{R}_{\pi/D} \equiv \frac{\mathcal{B}(\bar B^0 \to D_s^{*-}
\pi^+)}{\mathcal{B}(\bar B^0 \to D_s^{*-}D^{+} )}=\left|\frac
{V_{ub} }{V_{cb}}\right|^2\left|\frac
{T_{D_s^*\pi}}{T_{D_s^*D}+P_{D_s^*D}}\right|^2
\left(\frac{p_c^\pi}{p_c^D}\right)=\left|\frac {V_{ub}
}{V_{cb}}\right|^2\times(0.45\pm 0.07),
\end{eqnarray}
where $p_c^X$ is the c.m. momentum of the decay particle $X$.
Combining the experimental data of Table 1, we arrive at the following ratios:
\begin{eqnarray}\label{eq:ratio6}
&&R_{\pi/D}=(0.32\pm 0.07)\times 10^{-2};\\
&&R_{\pi/D^*}=(0.29\pm 0.07)\times 10^{-2};\\
&&\widetilde{R}_{\pi/D}=(0.34\pm 0.10)\times 10^{-2}.
\end{eqnarray}
After compared  with the predictions of the PQCD approach (\ref{eq:ratio1}-\ref{eq:ratio5}),
we can get:
\begin{eqnarray}\label{eq:ratio7}
\left|\frac {V_{ub}
}{V_{cb}}\right|=\left\{
                   \begin{array}{ll}
                     0.085\pm0.011 & \mathrm{from}~~~{R_{\pi/D}~,} \\
                     0.079\pm0.011 & \mathrm{from}~~~{R_{\pi/D^*}~,} \\
                     0.087\pm0.014 & \mathrm{from} ~~~{\widetilde{R}_{\pi/D}~,}
                   \end{array}
                 \right.
\end{eqnarray}
and the averaged value is
\begin{eqnarray}\label{ave}
\left|\frac {V_{ub}}{V_{cb}}\right|=0.083\pm0.007.
\end{eqnarray}
If we set $|V_{cb}|=0. 0415\pm0.0011$ which is the global fit PDG result, we can get
\begin{eqnarray}
\left|{V_{ub}}\right|=(3.44\pm 0. 30)\times 10^{-3}.
\end{eqnarray}
Compared with the PDG values Eqs. (\ref{vubexp1}--\ref{vubexp3}),
our central value favors a bit smaller $|V_{ub}|$, but agrees well with the exclusive data.
We also notice that the uncertainty is still large, which is because there are larger uncertainties
in both theoretical calculations and experimental data.
With much more knowledge in QCD and precisely measured experimental data in future, the errors in our prediction
will be reduced.

With the averaged ratio of $\left|V_{ub}/V_{cb}\right|$ Eq. (\ref{ave}), by using both a relation (\ref{eq:ratio2})
and experimental data for $\bar{B}^0 \to D_s^-D^+$, we can predict
the branching ratio of  $\bar B^0 \to D_s^-\rho^+$  as:
\begin{eqnarray}\label{brratio1}
{\mathcal{B}(\bar B^0 \to D_s^-\rho^+)}=(2.7\pm 1.2)\times 10^{-5},
\end{eqnarray}
which would be  measured easily in future $B$ factories.  Please also note that our theoretical estimate
within the PQCD approach for the branching ratio, as shown in Table 1, is
\begin{eqnarray}
{\mathcal{B}(\bar B^0 \to D_s^-\rho^+)}=  (3.1^{+1.0+1.3}_{-0.8-1.0})\times 10^{-5}.
\end{eqnarray}
To our best knowledge
about this channel, still there are only upper limits from  CLEO \cite{Alexander:1993gp} and ARGUS
\cite{Albrecht:1993ce}. We hope future $B$ factories could measure this channel to test our prediction.

Finally we add one comment on $B_s$ decay: With the help from  CDF, $\mathrm{D}{\O}$ and LHC,
the measurement of branching ratios of $\bar B_s^0$ decay modes becomes much more precise than before.
In principle, we can generalize the strategy of $\bar B_d$ decays to $\bar B_s$ by using SU(3) symmetry.
We can define the ratio similarly:
\begin{eqnarray}
R_{K/D}^s \equiv \frac{\mathcal{B}(\bar B_s^0 \to D_s^-
K^+)}{\mathcal{B}(\bar B_s^0 \to D_s^-D_s^+ )}~.
\end{eqnarray}
However, both decay modes $\bar B_s^0 \to D_s^-
K^+$ and $\bar B_s^0 \to D_s^-D_s^+ $ are much more complicate than  decay modes of $\bar B_d^0$.
Primarily because both $B_s^0$ and  $\bar B_s^0$ can decay to the same final state $D_s^- K^+$,
we cannot discriminate whether the final states come from $B_s^0$ or $\bar B_s^0$.
The same situation occurs in  $\bar B_s^0 \to D_s^-D_s^+ $ decay mode.
Furthermore, the $\bar B_s^0 \to D_s^- K^+$ mode also has the annihilation diagrams, which
will lead to larger uncertainty in calculating the $R_{K/D}^s$.
In practice, in order to measure the CKM angle $\gamma$,
the experimentalists often analyze the time dependent CP violation of these kinds of decay modes.

As summary, we revisited extracting  $\left|V_{ub}/V_{cb}\right|$ and
$|V_{ub}|$ through calculating the ratios   ${\cal B}(B^0\to D_s^{(*)+}(\pi^-,\rho^-)/
{\cal B}(B^0\to D_s^{(*)+} D^{(*)-})$ in the PQCD approach. Using the updated distribution
amplitudes and the latest experimental data, we got  $\left|{V_{ub}}/{V_{cb}}\right|=0.083\pm0.007$,
which favors a bit smaller $|V_{ub}|$ compared to the value of PDG.
Moreover, we predicted the branching ratio of $\bar B^0 \to D_s^-\rho^+$ $\sim$ $(2.7\pm 1.2)\times 10^{-5}$,
which can be measured in future $B$ factories. Furthermore,
we could test the applicability of the PQCD with charmed mesons through comparing the future data and our calculation.


\section*{Acknowledgement}
We would like to thank Y.J. Kwon for his valuable comments.
The work of C.S.K. is supported in part by Basic Science Research Program through the NRF of Korea
funded by MOEST (2009-0088395),  in part by KOSEF through the Joint Research Program (F01-2009-000-10031-0),
and in part by WPI Initiative, MEXT, Japan.
The work of Y.L. was supported by the Brain Korea 21
Project and by the National Science Foundation under contract
Nos.10805037 and 10625525.


\newpage

\begin{thebibliography}{99}
\bibitem{Isidori:2008qp}
  G.~Isidori,
  arXiv:0801.3039 [hep-ph].



\bibitem{Charles:2004jd}
  J.~Charles {\it et al.}  [CKMfitter Group],
  Eur.\ Phys.\ J.\  C {\bf 41}, 1 (2005)
  [arXiv:hep-ph/0406184].

\bibitem{Lee:2008vs}
  K.~S.~M.~Lee,
  Phys.\ Rev.\  D {\bf 78}, 013002 (2008)
  [arXiv:0802.0873 [hep-ph]].

\bibitem{Lange:2005yw}
  B.~O.~Lange, M.~Neubert and G.~Paz,
  Phys.\ Rev.\  D {\bf 72}, 073006 (2005)
  [arXiv:hep-ph/0504071].


\bibitem{Dalgic:2006dt}
  E.~Dalgic, A.~Gray, M.~Wingate, C.~T.~H.~Davies, G.~P.~Lepage and J.~Shigemitsu,
  Phys.\ Rev.\  D {\bf 73}, 074502 (2006)
  [Erratum-ibid.\  D {\bf 75}, 119906 (2007)]
  [arXiv:hep-lat/0601021].

\bibitem{Okamoto:2004xg}
  M.~Okamoto {\it et al.},
  Nucl.\ Phys.\ Proc.\ Suppl.\  {\bf 140}, 461 (2005)
  [arXiv:hep-lat/0409116].

\bibitem{Ball:2004ye}
  P.~Ball and R.~Zwicky,
  Phys.\ Rev.\  D {\bf 71}, 014015 (2005)
  [arXiv:hep-ph/0406232].

\bibitem{Chen:2009qk}
  C.~H.~Chen, Y.~L.~Shen and W.~Wang,
  Phys.\ Lett.\  B {\bf 686}, 118 (2010)
  [arXiv:0911.2875 [hep-ph]].


\bibitem{Amsler:2008zzb}
 C.~Amsler {\it et al.}  [Particle Data Group],
  Phys.\ Lett.\  B {\bf 667}, 1 (2008)(URL: http://pdg.lbl.gov).

\bibitem{Koide:1989hv}
  Y.~Koide,
  Phys.\ Rev.\  D {\bf 39}, 3500 (1989);
D.~Choudhury, D.~Indumati, A.~Soni and S.~Uma Sankar,
  Phys.\ Rev.\  D {\bf 45}, 217 (1992).


\bibitem{Kim:2000cv}
  C.~S.~Kim, Y.~Kwon, J.~Lee and W.~Namgung,
  Phys.\ Rev.\  D {\bf 63}, 094506 (2001)
  [arXiv:hep-ph/0010157].

\bibitem{Beneke:1999br}
  M.~Beneke, G.~Buchalla, M.~Neubert and C.~T.~Sachrajda,
  Phys.\ Rev.\ Lett.\  {\bf 83}, 1914 (1999)
  [arXiv:hep-ph/9905312];\\
 M.~Beneke, G.~Buchalla, M.~Neubert and C.~T.~Sachrajda,
  Nucl.\ Phys.\  B {\bf 591}, 313 (2000)
  [arXiv:hep-ph/0006124].

\bibitem{Keum:2000ph1}
  Y.~Y.~Keum, H.~n.~Li and A.~I.~Sanda,
  Phys.\ Lett.\  B {\bf 504}, 6 (2001)
  [arXiv:hep-ph/0004004].

\bibitem{Keum:2000ph}
Y.~Y.~Keum, H.~N.~Li and A.~I.~Sanda,
  Phys.\ Rev.\  D {\bf 63}, 054008 (2001)
  [arXiv:hep-ph/0004173];\\
  C.~D.~Lu, K.~Ukai and M.~Z.~Yang,
  Phys.\ Rev.\  D {\bf 63}, 074009 (2001)
  [arXiv:hep-ph/0004213];\\
   A.~Ali, G.~Kramer, Y.~Li, C.~D.~Lu, Y.~L.~Shen, W.~Wang and Y.~M.~Wang,
  Phys.\ Rev.\  D {\bf 76}, 074018 (2007)
  [arXiv:hep-ph/0703162].

\bibitem{Bauer:2000yr}
  C.~W.~Bauer, S.~Fleming, D.~Pirjol and I.~W.~Stewart,
  Phys.\ Rev.\  D {\bf 63}, 114020 (2001)
  [arXiv:hep-ph/0011336];\\
 C.~W.~Bauer and I.~W.~Stewart,
  Phys.\ Lett.\  B {\bf 516}, 134 (2001)
  [arXiv:hep-ph/0107001].

\bibitem{Keum:2003js}
  Y.~Y.~Keum, T.~Kurimoto, H.~N.~Li, C.~D.~Lu and A.~I.~Sanda,
  Phys.\ Rev.\  D {\bf 69}, 094018 (2004)
  [arXiv:hep-ph/0305335].


\bibitem{Li:2008ts}
  R.~H.~Li, C.~D.~Lu and H.~Zou,
  Phys.\ Rev.\  D {\bf 78}, 014018 (2008)
  [arXiv:0803.1073 [hep-ph]].

\bibitem{Zou:2009zza}
  H.~Zou, R.~H.~Li, X.~X.~Wang and C.~D.~Lu,
  J.\ Phys.\ G {\bf 37}, 015002 (2010)
  [arXiv:0908.1856 [hep-ph]].

\bibitem{Li:2009xf}
  R.~H.~Li, X.~X.~Wang, A.~I.~Sanda and C.~D.~Lu,
  Phys.\ Rev.\  D {\bf 81}, 034006 (2010)
  [arXiv:0910.1424 [hep-ph]].

\bibitem{Buchalla:1995vs}
   For a review, see G.~Buchalla, A.~J.~Buras and M.~E.~Lautenbacher,
  Rev.\ Mod.\ Phys.\  {\bf 68}, 1125 (1996)
  [arXiv:hep-ph/9512380].

\bibitem{Ball:2006wn}
  P.~Ball, V.~M.~Braun and A.~Lenz,
  JHEP {\bf 0605}, 004 (2006)
  [arXiv:hep-ph/0603063].\\
   P.~Ball, V.~M.~Braun and A.~Lenz,
  JHEP {\bf 0708}, 090 (2007)
  [arXiv:0707.1201 [hep-ph]].

\bibitem{Grozin:1996pq}
  A.~G.~Grozin and M.~Neubert,
  Phys.\ Rev.\  D {\bf 55}, 272 (1997)
  [arXiv:hep-ph/9607366];\\
  H.~Kawamura, J.~Kodaira, C.~F.~Qiao and K.~Tanaka,
  Phys.\ Lett.\  B {\bf 523}, 111 (2001)
  [arXiv:hep-ph/0109181].

\bibitem{utfitter}
M. Bona et al. [UTfit Collaboration], JHEP 0803, 049 (2008) [arXiv:0707.0636].

\bibitem{Alexander:1993gp}
  J.~P.~Alexander {\it et al.}  [CLEO Collaboration],
  Phys.\ Lett.\  B {\bf 319}, 365 (1993).

\bibitem{Albrecht:1993ce}
  H.~Albrecht {\it et al.}  [ARGUS Collaboration],
  Z.\ Phys.\  C {\bf 60}, 11 (1993).

\end{thebibliography}
\end{document}